%% file: arxiv.tex
\documentclass[letterpaper]{article} % DO NOT CHANGE THIS
\usepackage[preprint]{aaai2027}  % DO NOT CHANGE THIS
\usepackage[hyphens]{url}  % DO NOT CHANGE THIS
\usepackage{graphicx} % DO NOT CHANGE THIS
\usepackage{natbib}  % DO NOT CHANGE THIS AND DO NOT ADD ANY OPTIONS TO IT
\usepackage{caption} % DO NOT CHANGE THIS AND DO NOT ADD ANY OPTIONS TO IT
\usepackage{algorithm}
\usepackage{algorithmic}

\usepackage{newfloat}
\usepackage{listings}
\DeclareCaptionStyle{ruled}{labelfont=normalfont,labelsep=colon,strut=off} % DO NOT CHANGE THIS
\floatstyle{ruled}
\newfloat{listing}{tb}{lst}{}
\floatname{listing}{Listing}

\usepackage{booktabs}

\usepackage{lipsum}
\usepackage{multirow}
\usepackage[table]{xcolor}
\newcommand\zty[1]{{\color{black}#1}}
\newcommand\cj[1]{{\color{black}#1}}

\title{CryptoFlowBench: Benchmarking Multi-Hop Reasoning on Blockchains for Cryptocurrency Fund Tracing}
\author{
    Runang He\textsuperscript{\rm 1},
    Tongya Zheng\textsuperscript{\rm 1,2,3}\thanks{Corresponding Author},
    Jing Cao\textsuperscript{\rm 1,2,3},
    Yucheng Ying\textsuperscript{\rm 1},
    Shunyu Liu\textsuperscript{\rm 1},
    Bingde Hu\textsuperscript{\rm 1},
    Yang Gao\textsuperscript{\rm 1},
    Zeyuan Hu\textsuperscript{\rm 4},
    Jianrong Zhang\textsuperscript{\rm 5},
    Mingli Song\textsuperscript{\rm 1,2,3}
}
\affiliations{
    \textsuperscript{\rm 1}State Key Laboratory of Blockchain and Data Security, Zhejiang University\\
    \textsuperscript{\rm 2}Hangzhou City University\\
    \textsuperscript{\rm 3}Hangzhou High-Tech Zone (Binjiang) Institute of Blockchain and Data Security\\
    \textsuperscript{\rm 4}China Mobile Group Zhejiang Co., Ltd\\
    \textsuperscript{\rm 5}China Unicom Digital Technology Co., Ltd\\
    herunang@zju.edu.cn, 
    doujiang\_zheng@163.com, 
    jingcao@hzcu.edu.cn,
    yingyu@zju.edu.cn,
    shunyu.liu.cs@gmail.com,
    tonyhu@zju.edu.cn,
    roygao@zju.edu.cn,
    huzeyuan@zj.chinamobile.com,
    zhangjr67@chinaunicom.cn,
    brooksong@zju.edu.cn
}

\begin{document}

\maketitle

\begin{abstract}
\zty{
% The pseudonymous accounts and abundant smart contracts on blockchains have attracted massive illicit activities, severely hindering tracing cryptocurrency fund flows of suspicious groups.
% On the one hand, advanced Graph Neural Networks have shown remarkable progress on fraud detection on blockchains, yet one-hop detection can hardly generalize to multi-hop reasoning on fund networks.
% On the other hand, recent fund flow tracing methods mostly design specialized heuristics to capture multi-hop relationships rather than intelligent multi-hop reasoning.
% Therefore, we are motivated to curate a cryptocurrency fund benchmark based on authoritative reports of illicit incidents, called CryptoFundBench, to comprehensively evaluate the effectiveness of various baseline methods on node-level, edge-level, and multi-hop tasks.
% Moreover, we design a flexible yet powerful agent to complete tasks of the proposed benchmark, called ChainSleuth, which simply conducts a ReAct-loop with graph tools until termination.
% Experiments on four tasks at different levels have demonstrated that no baseline methods can dominate all four tasks since they are particularly designed for one of these tasks.
% This observation validates the significance of CryptoFlowBench that researchers should improve the line of tasks instead of focusing on only a single task.
% Our proposed ChainSleuth has obtained noteworthy gains on multi-hop reasoning tasks, indicating the superiority of agentic reasoning on graphs.
Pseudonymous accounts and complex smart-contract ecosystems substantially hinder the tracing of illicit cryptocurrency fund flows. 
Existing Graph Neural Networks (GNNs) have achieved promising performance in blockchain fraud detection, yet their static node- or edge-level predictions do not naturally generalize to multi-hop fund tracing. 
Meanwhile, existing tracing methods mainly rely on predefined heuristics to capture multi-hop relationships, limiting their adaptive reasoning capabilities. 
% To bridge this gap, we introduce CryptoFlowBench, a benchmark constructed from authoritative reports of cryptocurrency-related illicit incidents, with four tasks spanning node-, edge-, and multi-hop fund analysis.
% This comprehensive benchmark poses significant tracing challenges to existing heuristic-based methods.
% Therefore, we further propose ChainSleuth, a graph-tool-augmented agent that iteratively reasons and acts through a ReAct loop until termination, which enables agent to acquire on-chain information adaptively and recover from wrong path.
To bridge this gap, we introduce CryptoFlowBench, a benchmark constructed from authoritative reports on cryptocurrency-related illicit incidents, with four tasks spanning local to multi-hop fund tracing. 
These tasks pose substantial challenges to existing learning- and heuristic-based methods, particularly in complex multi-hop scenarios. 
We therefore propose ChainSleuth, a graph-tool-augmented agent that iteratively reasons, invokes graph tools, and updates its tracing trajectory, enabling adaptive information acquisition and recovery from erroneous paths.
Experiments show that no baseline consistently dominates all four tasks, revealing substantial gaps between local fraud detection and multi-hop fund tracing. 
ChainSleuth achieves notable improvements on multi-hop tasks, demonstrating the potential of agentic graph reasoning for cryptocurrency fund investigation.
}
\end{abstract}

\begin{links}
    \link{Code}{https://github.com/LuoXishuang0712/CryptoFlowBench}
    % \link{Datasets}{https://aaai.org/example/datasets}
    % \link{Extended version}{https://aaai.org/example/extended-version}
\end{links}

\section{Introduction}

\begin{figure}[ht!]
    \centering
    \includegraphics[width=\linewidth]{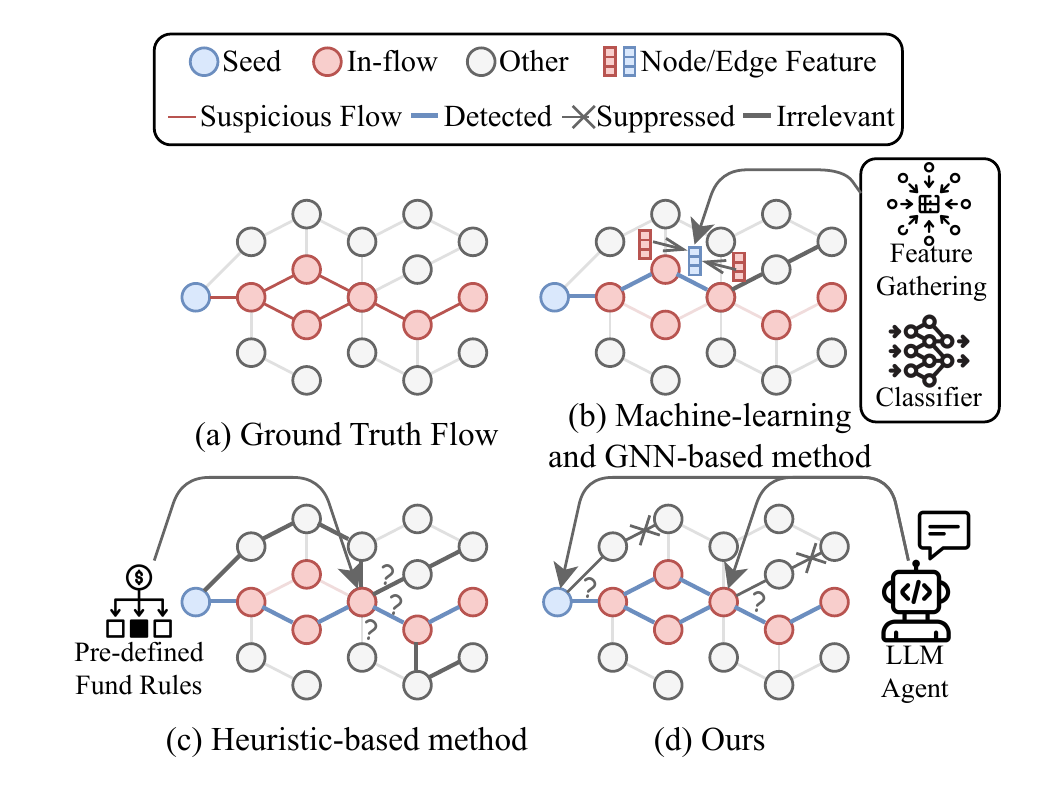}
    \caption{
    Comparison of fund-tracing behaviors on a case-centered transaction graph.
    Red edges in (a) show the verified suspicious flow.
    The machine-learning/GNN-based method in (b) cannot recover after selecting an irrelevant branch, shown as gray edges, while the rule-based method in (c) over-expands under a fixed propagation policy.
    ChainSleuth in (d) retrieves on-chain evidence, revises earlier decisions, and suppresses unsupported expansions.
    }
    \label{fig:trace_example}
\end{figure}

Public blockchains provide transparent and append-only transaction records without directly binding accounts to real-world identities.
While supporting digital payments and decentralized finance, this infrastructure can also be exploited to steal, conceal, and launder illicit funds.
According to the FBI's 2025 Internet Crime Report, cryptocurrency was involved in approximately \$11 billion of nearly \$20 billion in reported losses~\cite{fbi2025internetcrime}.
To analyze such illicit activities, on-chain transfers can be represented as large, temporal, and attributed money-flow graphs (MFGs), where nodes correspond to addresses or contracts and directed edges represent asset transfers~\cite{turner2020analysis,qi2023blockchain}.
Within these graphs, some nodes are associated with services such as decentralized exchanges and cross-chain bridges, introducing heterogeneous transaction patterns into the tracing process~\cite{liu2021knowledge,bartoletti2021cryptocurrency,cunjak2022crypto,almeida2023review}.
% Practical investigation therefore requires more than identifying suspicious addresses or transactions: investigators must reconstruct multi-hop fund flows, select relations for further examination, and stop when evidence no longer supports continued expansion.
% Fund tracing is thus a sequential, evidence-sensitive graph investigation problem rather than an isolated classification task.
As shown in Figure~\ref{fig:trace_example}(a), practical fund tracing aims to recover the suspicious flow from limited case evidence while excluding irrelevant transfers.
Existing fund-tracing methods broadly follow two paradigms.
Heuristic approaches propagate funds using deterministic taint-allocation rules or graph-based relevance and flow objectives~\cite{moser2014towards,tironsakkul2019probing,wu2023tracer,lin2024denseflow}.
Learning-based approaches use classical classifiers or GNNs to prioritize suspicious transactions and addresses from structural and behavioral features~\cite{weber2019anti,shen2021identity,huang2024peae}.
However, fixed propagation policies and independent local predictions provide limited support for revising earlier decisions, while practical tracing requires evidence-driven multi-hop reasoning and adaptive stopping.

Supporting and evaluating such practical tracing requires addressing three key challenges.
\textbf{Challenge 1: Case-grounded evaluation.}
Large detection benchmarks mainly provide binary address or transaction labels, whereas tracing studies often use independently constructed case graphs with inconsistent scopes, labels, and protocols.
Neither setting supports unified evaluation of complete investigations across heterogeneous incidents.
\textbf{Challenge 2: Error accumulation.}
Machine-learning and GNN-based methods generally make independent local predictions, allowing an early false positive to propagate through a multi-hop trajectory without a mechanism for later revision, \cj{causing error accumulated along the propagation path}, as illustrated in Figure~\ref{fig:trace_example}(b).
\textbf{Challenge 3: Adaptive exploration and boundary control.}
Heuristic-based tracing methods rely on predefined attribution or structural policies~\cite{moser2014towards,tironsakkul2019probing,wu2023tracer,lin2024denseflow,wu5076472shadows}.
Such policies may expand through irrelevant but highly connected services, causing common on-chain service nodes to repeatedly divert the tracing process, particularly when no reliable seed address is available, as shown in Figure~\ref{fig:trace_example}(c).

To address \textbf{Challenge 1} and make the sequential and boundary-control requirements in \textbf{Challenges 2 and 3} measurable, we construct \textsc{CryptoFlowBench}, a case-grounded benchmark that aligns heterogeneous incident reports with locally verified transactions.
Each incident is represented as an evidence graph containing address roles, tracing actions, path stages, hard negatives, evidential boundaries, and source provenance.
We deterministically derive four complementary tasks under a unified protocol:
\textbf{transaction existence} for factual verification;
\textbf{next-step action} for seeded decisions among \emph{follow}, \emph{inspect}, \emph{stop}, and \emph{ignore};
\textbf{path completion} for multi-hop tracing under accumulated errors; and
\textbf{subgraph filtering} for seed localization and tracing in distractor-rich graphs.
These tasks separate factual grounding, local decisions, sequential reconstruction, and seedless investigation, while exposing over-expansion and boundary violations.

To address \textbf{Challenges 2 and 3}, we propose \textsc{ChainSleuth}, an evidence-grounded, tool-augmented LLM agent.
Within a bounded ReAct loop~\cite{yao2022react}, ChainSleuth maintains an evolving investigation state, selects explicit tracing actions, and retrieves transaction-level evidence through external tools.
New observations can be used to revisit earlier hypotheses, suppress unsupported branches, and stop at evidential boundaries, as illustrated in Figure~\ref{fig:trace_example}(d).
Experiments show that strong local prediction does not necessarily translate into effective end-to-end tracing.
Although conventional classifiers remain competitive on single-step decisions, ChainSleuth performs better on \textit{Path} and \textit{Subgraph} tasks, improving Path F1 by 70.0\%, reducing over-expansion by 26.6\%, and improving predicted-seed F1 by 21.4\% over the strongest baseline for each metric.

Our main contributions are summarized as follows:
\begin{itemize}
    \item \textbf{CryptoFlowBench: a case-grounded benchmark for Ethereum fund-flow investigation.}
    We align heterogeneous incident reports with locally verified transactions to construct case-centered evidence graphs that summarizing on-chain incidents.
    Based on 22 incidents across five types, the benchmark derives four complementary tasks under a unified protocol, covering factual verification, seeded local decisions, multi-hop path reconstruction, and seedless suspicious-region discovery.

    \item \textbf{ChainSleuth: an evidence-grounded, tool-augmented LLM agent.}
    ChainSleuth maintains an evolving investigation state, selects explicit tracing actions, and retrieves auditable on-chain evidence through external tools.
    It can revisit earlier hypotheses as new evidence is acquired, thereby correcting accumulated errors, suppressing unsupported branches, and controlling tracing boundaries.

    \item \textbf{A comprehensive, subtask-specific evaluation.}
    We compare ChainSleuth with task-appropriate heuristic-based, machine-learning, GNN, and LLM baselines under matched case splits and tracing budgets.
    % The evaluation separates local decision quality from multi-hop error accumulation, seed localization from path propagation, and tracing coverage from irrelevant expansion, while additionally examining tool-use and execution reliability.
    On the core tracing tasks, ChainSleuth improves Path F1 by 70.0\%, reduces over-expansion by 26.6\%, and improves predicted-seed F1 by 21.4\% over the strongest specialized baseline for each metric.
\end{itemize}

\section{Related Work}

% \subsection{Datasets and Learning Methods for Cryptocurrency Illicit-Activity Detection}
\subsection{Cryptocurrency Illicit-Activity Detection}

% Existing cryptocurrency-security datasets have progressively expanded illicit-activity detection from individual transactions to addresses and subgraphs. 
% Elliptic provides a temporal Bitcoin transaction graph for illicit transaction classification~\cite{weber2019anti}, while Elliptic++ augments it with more than 822,000 wallet addresses, address-level features, and multiple Address-Transaction graph representations, enabling both transaction- and address-level detection~\cite{elmougy2023demystifying}. 
% EthereumHeist organizes asset flows from real Ethereum thefts into laundering stages and behavioral patterns~\cite{wu2023toward}, and Elliptic2 formulates money laundering as subgraph classification over a large collection of labeled Bitcoin transaction subgraphs~\cite{bellei2024shape}. 
% These datasets broaden illicit-activity detection from transactions to addresses and subgraphs, but do not capture the operational process of reconstructing an incident.
Existing cryptocurrency illicit-activity detection methods can be broadly grouped into several paradigms~\cite{fouzy2026cryptocurrency}.
Heuristic-based methods include rule-based approaches using predefined attribution policies and graph-based approaches using transaction topology for candidate prioritization; although interpretable, they are limited by fixed assumptions or scalability on large, densely connected graphs.
Machine-learning methods, such as logistic regression, random forests, and XGBoost~\cite{weber2019anti,elmougy2023demystifying}, are generally efficient but depend heavily on handcrafted features and capture relational dependencies only indirectly.
GNN-based methods, including GCN~\cite{weber2019anti}, I$^2$BGNN~\cite{shen2021identity}, and PEAE-GNN~\cite{huang2024peae}, model complex account interactions more directly, but their computational cost and limited explainability hinder evidence-sensitive forensic use.
General graph anomaly detection methods, such as H$_2$GCN~\cite{zhu2020beyond}, PMP~\cite{zhuo2024partitioning}, ConsisGAD~\cite{chen2024consistency}, UNPrompt~\cite{niu2024zero}, ARC~\cite{liu2024arc}, and AnomalyGFM~\cite{qiao2025anomalygfm}, are not blockchain-specific but offer mechanisms for handling heterophily, label scarcity, and class imbalance.
Their effectiveness has also been demonstrated in blockchain anomaly detection~\cite{he2026temporal}, although they remain focused on node-level detection rather than evidence-grounded, multi-hop fund tracing.

\subsection{Cryptocurrency Fund-Flow Tracing}

Early cryptocurrency tracing relied on deterministic taint-propagation rules, including Poison, Haircut, FIFO, LIFO, and TIHO, to associate downstream Bitcoin outputs with known illicit funds~\cite{moser2014towards,tironsakkul2019probing}. 
Although transparent and reproducible, these methods depend on fixed value-allocation assumptions and can produce substantially different tracing results for the same source transaction. 
Learning- and graph-based approaches subsequently introduced more flexible prioritization. 
The Elliptic study applied classical classifiers and graph convolutional network (GCN) to illicit Bitcoin transaction detection~\cite{weber2019anti}, while TRacer modeled tracing in account-based blockchains as relevance-guided search over directed, temporal, weighted, and multi-relational graphs~\cite{wu2023tracer}. 
DenseFlow and DenseFlow+ further combined suspicious-group discovery with Ethereum asset-flow reconstruction using dense-subgraph and flow-based objectives~\cite{lin2024denseflow,wu5076472shadows}. 
These methods advance transaction classification, account prioritization, and path reconstruction, but do not explicitly model evidence-grounded investigative decisions or revision after new observations.

\subsection{LLM Agents for Blockchain Security and Analysis}

LLM agents extend static language-model generation by interleaving reasoning, tool use, and environmental observations, enabling adaptive multi-step decision making~\cite{yao2022react, paranjape2023art}. Recent studies have applied this agentic paradigm to external environments such as web navigation~\cite{zhou2024webarena, lu2025deepdive}, graphical user interfaces~\cite{xie2024osworld,zhai2025agentevolver}, and structured graphs~\cite{jin2024graphcot,ying2026graphscout}, where agents iteratively gather evidence and refine their actions. Such capabilities are well suited to on-chain analysis, which requires querying large transaction graphs and verifying intermediate decisions with external evidence. LLM agents have also been explored for smart-contract auditing, vulnerability detection, program analysis, and automated repair~\cite{he2024large,wei2024llm}. However, existing studies largely focus on contract code rather than evidence-grounded, multi-step fund-flow tracing.
% Retrieval-augmented generation (RAG) grounds language-model outputs in external evidence~\cite{lewis2020retrieval}, while ReAct-style agents interleave reasoning, tool use, and environmental observations to support adaptive multi-step decision making~\cite{yao2022react}. 
% These capabilities are well suited to on-chain investigation, where transaction graphs are too large to place directly in context and factual decisions must be verified through external queries. 
% LLMs and multi-agent systems have recently been applied to blockchain security, particularly smart-contract auditing, vulnerability detection, program analysis, and repair~\cite{he2024large,wei2024llm}. 
% These systems demonstrate the potential of LLM agents in blockchain security, but do not address evidence-grounded, multi-step transaction tracing.

\section{Preliminaries}
\label{sec:preliminaries}

\paragraph{Blockchain Ledger}

An account-based blockchain is an append-only block sequence
$\mathcal{L}=(B_0,\ldots,B_L)$.
Within a block window $W=[b_{\min},b_{\max}]$, each indexed transfer is represented as
\[
x_i=(h_i,b_i,t_i,s_i,d_i,a_i,m_i,\kappa_i)\in\mathcal{X}_W,
\]
where $h_i$, $b_i$, and $t_i$ denote the transaction hash, block number, and timestamp;
$s_i$ and $d_i$ denote the source and destination addresses; and
$a_i$, $m_i$, and $\kappa_i$ denote the asset identifier, amount, and transfer type.
The indexed transfer set is
\[
\mathcal{X}_W=\{x_i\mid b_{\min}\leq b_i\leq b_{\max}\}.
\]
Exact transaction-existence claims are made only with respect to the indexed chain, block window, and supported transfer types.

\paragraph{Transaction-Graph Construction}

Raw transfers are organized into a directed address-relation graph whose nodes are addresses and whose relation edges retain the underlying transfer records.
For each ordered address pair $(u,v)$, define
\[
\mathcal{X}_{uv}=\{x_i\in\mathcal{X}_W\mid s_i=u,\ d_i=v\}.
\]
When $\mathcal{X}_{uv}\neq\emptyset$,
$e_{uv}=\mathrm{Agg}(\mathcal{X}_{uv})$ preserves transaction identifiers and transfer metadata, with amounts aggregated only among records sharing the same asset identifier and transfer type.
The resulting graph is
\[
\Phi_W(\mathcal{L})=\mathcal{G}_W=(V_W,E_W),
\]
where
\[
V_W=\{s_i,d_i\mid x_i\in\mathcal{X}_W\},
\qquad
E_W=\{e_{uv}\mid\mathcal{X}_{uv}\neq\emptyset\}.
\]

\paragraph{Fund-Flow Tracing}

Given $\mathcal{G}_W$, public incident evidence $x_c^{\mathrm{pub}}$, an optional seed set $S_0$, and exploration budget $\mathbf{B}$, fund-flow tracing returns
\[
\mathrm{Trace}(\mathcal{G}_W,x_c^{\mathrm{pub}},S_0;\mathbf{B})
=(\widehat{S},\widehat{E},\widehat{\mathcal{P}}),
\]
where $\widehat{S}$ is the supplied or inferred seed set,
$\widehat{E}$ is the selected tracing subgraph, and
$\widehat{\mathcal{P}}$ contains directed paths whose edges are contained in $\widehat{E}$.
For a hidden verified reference graph
$\mathcal{G}_c^*=(V_c^*,E_c^*)$, the objective is to recover relevant edges under $\mathbf{B}$ while limiting irrelevant expansion $\widehat{E}\setminus E_c^*$.
The reference graph represents the verified evidence for case $c$, rather than an exhaustive reconstruction of all illicit activity.

% \section{ Method}
% \label{sec:method}
\section{CryptoFlowBench}

\begin{table}[t!]
    \centering
    \scalebox{0.8}{
        \input{tables/task_info}
    }
    \caption{Incident types and primary sources of the benchmark cases.}
    \label{tab:task_info}
\end{table}

\begin{figure*}[ht!]
    \centering
    \includegraphics[width=\textwidth]{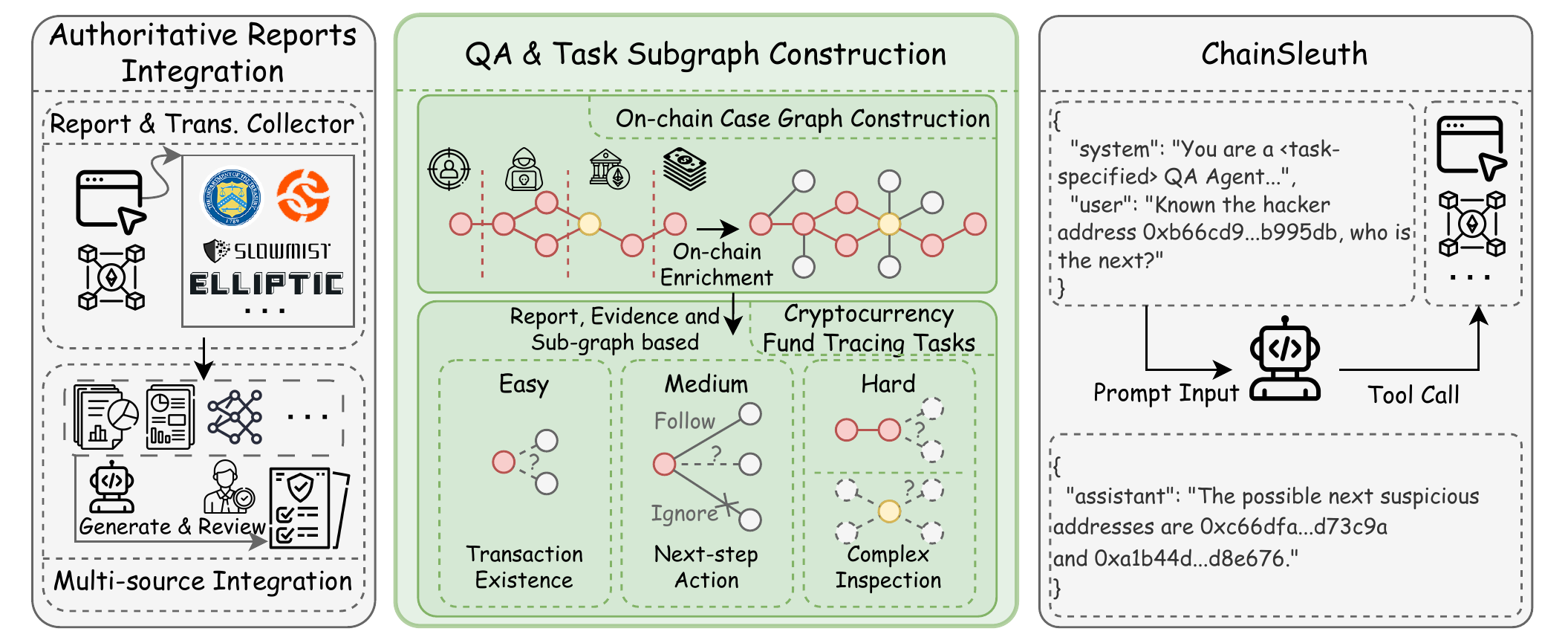}
    \caption{Overview of the benchmark construction pipeline.
    Public incident reports are consolidated into structured case summaries,
    aligned with indexed on-chain transactions, and converted into grounded question-answering tasks with task-specific verification. Where ``Trans.'' indicates ``Transaction''.}
    \label{fig:method_framework}
\end{figure*}

We propose CryptoFlowBench, a benchmark with four complementary fund-tracing tasks: transaction existence, next-step action, path completion, and subgraph filtering.
They evaluate factual verification, local tracing decisions, sequential reasoning under accumulated errors, and seedless subgraph-level tracing, respectively.

Figure~\ref{fig:method_framework} summarizes the benchmark construction pipeline.
Let $\mathcal{C}=\{c_i\}_{i=1}^{N}$ denote publicly documented cryptocurrency incidents.
We use LLM-driven agents
$\mathcal{A}_{\mathrm{col}}$ and $\mathcal{A}_{\mathrm{sum}}$
for report collection and structured summarization, while graph alignment, task generation, and verification are deterministic.
For each incident $c$, the agents collect a document set $\mathcal{D}_c$ and normalize its claims into a structured summary $S_c$.
Subsequent stages construct a report-evidence graph $\mathcal{H}_c$, align it with a bounded on-chain graph $\mathcal{G}_c^{(k)}$, and generate task instances $\mathcal{Q}_c$.
% The benchmark exposes a tool-grounded evaluation profile and a context-only profile used for ablation, while keeping oracle annotations separate from all public agent inputs.
According to this pipeline, we collect 22 incidents across five incident types, as shown in Table~\ref{tab:task_info}.

% \subsection{Report Collection}
\subsection{Authoritative Reports Integration}

Existing cryptocurrency datasets mainly support binary detection and provide limited case context that can be grounded in on-chain evidence.
This mismatch hinders the evaluation of practical tracing systems, especially LLM agents that require incident context and verifiable intermediate evidence.
We therefore design an authoritative report collection and integration pipeline.

We first collect several on-chain incidents and utilize a web-enabled LLM agent to collect online reports.
For each incident $c$, the collection agent produces
\[
\mathcal{D}_c=\mathcal{A}_{\mathrm{col}}(c;\theta_{\mathrm{col}},\mathcal{T}_{\mathrm{col}}),
\]
from public sources,
including incident disclosures, technical postmortems, security analyses, and blockchain explorer pages.
% The collection agent performs broad case-name searches followed by targeted page inspection and records the source URL, publisher, document type, and a coarse provenance-based source priority.
% We repeat collection with independently
% initialized search trajectories and merge deduplicated results to reduce retrieval variance.
To further reduce the randomness of the agent, we repeat the collection process for multiple passes, which is noted as $\mathcal{D}_c^{p}$, where $p$ indicates the index of pass.

Then, to verify and gather reports, we utilize a summarization agent to review all reports
\[
S_c=\mathcal{A}_{\mathrm{sum}}((\mathcal{D}_c^1, \ldots, \mathcal{D}_c^p);\theta_{\mathrm{sum}},\mathcal{T}_{\mathrm{sum}}).
\]
% The schema captures incident metadata, affected protocols and chains, reported losses, root causes, attacker and victim addresses, exploit transactions, and fund-flow descriptions,
% as well as recovery or sanction events when available.
% Each extracted claim retains references to its supporting sources.
% Missing or conflicting information is preserved as an open question rather than completed by unsupported model inference.
% This structured representation provides the report-level evidence used by the subsequent on-chain alignment stage.
The summarization agent independently verifies and distills the reports, and all sources are classified according to provenance priority.

% \subsection{On-chain Alignment and Case-Graph Construction}
\subsection{On-chain Case Graph Construction}

To align reports with on-chain evidence, we construct a case graph that maps report relations into on-chain transactions, ensuring all the evidence is case-grounded.

We first deterministically transform the structured summary $S_c$ into a case-evidence specification
\[
\mathcal{H}_c=(V_c^H,E_c^H,\rho_c,\pi_c),
\]
where $V_c^H$ contains reported addresses, contracts, protocols, and service entities;
$E_c^H$ contains reported transfers and stage-level relations;
$\rho_c$ records entity roles and tracing stages; and
$\pi_c$ preserves source provenance, temporal constraints, transaction identifiers, and verification status.
The evidence specification provides candidate anchors and alignment constraints, but does not itself assert that reported relations exist on chain.

Then, we ground and extend the report-derived evidence using the local blockchain index, which is constructed by fetching transactions from a blockchain client and indexing according to the addresses.
Given the max expand hop $k$, blockchain index $\mathcal{I}$, case evidence $\mathcal{H}_c$, case block window $W_c$, and resource bounds $\mathbf{B}$, we construct
\[
\mathcal{G}_c^{(k)}
=
\mathrm{Expand}_k(\mathcal{I},\mathcal{H}_c;W_c,\mathbf{B}).
\]
% The expansion operator first derives candidate anchors from explicitly reported attackers, victims, intermediaries, and transaction endpoints, and then retrieves their bounded $k$-hop neighborhoods from the local index.
% The hop parameter $k$ controls graph depth, while $\mathbf{B}$ limits the retained nodes, edges, and per-node neighbors.
% All observed edges originate from indexed transaction records and retain their transaction hashes, block numbers, assets, amounts, and transfer types.
During the expanding process, we first derive candidate anchors from explicitly reported entities, then retrieve their bounded $k$-hop neighborhoods from the local index.
% Reported relations in $E_c^H$ are aligned with indexed observations using their endpoints, direction, temporal constraints, transaction identifiers, and asset information.
% A reported relation is marked as chain-verified only when supported by a corresponding indexed observation.
Absence from the bounded expansion is not treated as evidence of non-existence; exact negative instances are generated only from complete index queries.
The resulting representation therefore combines report-derived semantics from $\mathcal{H}_c$ with locally verified transaction evidence from $\mathcal{G}_c^{(k)}$ through explicit alignment and provenance links.

% \subsection{Benchmark Task Construction}
\subsection{Cryptocurrency Fund Tracing Tasks}

Existing methods are typically evaluated on isolated objectives, such as node classification or fund-flow reconstruction, and therefore do not expose the full investigation process.
To better match practical requirements, CryptoFlowBench defines four complementary tasks: transaction existence, next-step action, path completion, and subgraph filtering.
% They are designed to evaluate the transaction query ability, next hop tracing decision quality, path extending ability, and practical trace mining ability, respectively.

First, given the aligned case evidence $\mathcal{H}_c$ and bounded on-chain graph $\mathcal{G}_c^{(k)}$, we jointly construct task instances and their oracle annotations through a deterministic, graph-first procedure.
Each candidate relation $e$ is assigned an evidence label $y_e\in\mathcal{Y}$ and an investigation action
\[
u_e\in\mathcal{U}
=
\{\mathtt{follow},\mathtt{inspect},\mathtt{stop},\mathtt{ignore}\},
\]
where \texttt{follow} denotes continuation along the reference fund flow, \texttt{inspect} marks relevant but insufficient evidence, \texttt{stop} identifies an evidential or operational boundary, and \texttt{ignore} denotes an irrelevant relation.
% The action set $\mathcal{U}$ is designed to transform task graph from binary edge classification to fund-flow tracing perspective, which pre-defines a soft border of the incidence in graph. 
This action space extends binary edge classification into an investigative decision interface, where \texttt{stop} and \texttt{ignore} explicitly represent the evidential boundary of a case.
% Raw transfers sharing the same ordered address pair are grouped into one candidate relation while retaining their transaction-level provenance.
% Candidate order is deterministically randomized to prevent positional leakage.

% From each labeled case graph, we generate four complementary task families:
Then we generate four types of tasks for each labeled case graph $\mathcal{G}_c^{(k)}$:
\textbf{transaction existence} (\textit{Trans}),
\textbf{next-step action} (\textit{Action}),
\textbf{path completion} (\textit{Path}), and
\textbf{subgraph filtering} (\textit{Subgraph}).
\textit{Trans} is a standalone diagnostic that asks whether an exact transaction relation exists under explicit block-window and transfer-type constraints.
% Its positive and negative labels are obtained only from complete index queries;
% absence from a sampled subgraph is never treated as evidence of non-existence.
\textit{Action} evaluates one seeded local decision over the outgoing relations of a known current node.
\textit{Path} composes the same local decision interface over multiple hops, exposing error accumulation under teacher-forced and free-rollout settings.
\textit{Subgraph} removes the trusted starting node and evaluates seed localization followed by tracing within a distractor-rich subgraph.

Let
\[
\mathcal{R}
=
\{\mathrm{Trans},\mathrm{Action},\mathrm{Path},\mathrm{Subgraph}\}.
\]
The task set for case $c$ is generated as
\[
\mathcal{Q}_c
=
\bigcup_{r\in\mathcal{R}}
\mathrm{Generate}_r
\bigl(
\mathcal{H}_c,
\mathcal{G}_c^{(k)};
\mathcal{Y},
\mathcal{U},
\mathbf{B}_r
\bigr),
\]
where $\mathbf{B}_r$ specifies task-dependent limits on depth, candidate size, node expansion, and evidence access.
Task \textit{Action}, \textit{Path}, and \textit{Subgraph} deliberately share the same edge-decision semantics, while \textit{Action} fixes the seed and evaluates one local state, \textit{Path} fixes the seed and evaluates a sequential rollout, and \textit{Subgraph} additionally requires the seed to be inferred.

Every task
$q\in\mathcal{Q}_c$
is separated into a public record
$x_q^{\mathrm{pub}}$
and a disjoint oracle record
$y_q^{\mathrm{oracle}}$.
The former contains only the task question, candidate set, permitted graph context, tool declarations, budget, and response schema.
Answers, gold actions, case-path labels, and evaluation annotations remain oracle-only and will be removed from prompts, retrieval stores, and tool outputs.
For a fixed case snapshot and configuration, candidate construction, ordering, oracle generation, and public-context sanitization are deterministic.

Appendix.A presents representative benchmark cases to illustrate how incident reports are aligned with on-chain evidence and converted into case-centered graphs with tracing labels, evidential boundaries, and task instances.

% The benchmark exposes two evaluation profiles.
% The primary \emph{tool-grounded} profile provides the minimum public state
% needed to define the task and requires the evaluated agent to retrieve or verify
% on-chain evidence through declared tools.
% A separately generated \emph{context-only} profile supplies a fixed public graph
% snapshot with tools disabled and is used only as a paired ablation.
% DTE remains a standalone capability diagnostic, whereas EAC, PC, and SNF form
% the main tracing benchmark.

% \subsection{ChainSleuth: On-chain Retrieval-Augmented Agent}
% \section{ChainSleuth: A ReAct-based Tracing Agent}
\section{ChainSleuth}

% Existing fund-flow tracing methods fail to be applied in practical tracing task, as they are designed to trace according to fixed amount rules, which highly rely on quality label, or are unable to correct circulated error.
% To address these problems, we propose ChainSleuth, a ReAct-based~\cite{yao2022react} and retrieval-augmented LLM agent to trace illicit fund-flow on-chain.
% By utilizing local blockchain index, Etherscan database, and pre-trained detectors as tools, ChainSleuth is able to retrieve chain evidence, and finally complete the fully trace.
% We will discuss the tool design in detail in Appendix.B.
Existing tracing methods rely on fixed propagation rules or independently scored local decisions, and therefore provide limited support for revising earlier hypotheses during multi-hop investigation.
We propose ChainSleuth, a ReAct-based, retrieval-augmented LLM agent for evidence-grounded on-chain fund tracing~\cite{yao2022react}.
Through a local blockchain index, blockchain explorers, and optional pretrained detectors, ChainSleuth adaptively retrieves evidence and updates its tracing trajectory.
The tool design is detailed in Appendix.B.

% We implement a retrieval-augmented LLM agent that treats the local Ethereum
% index as an external evidence environment.
For each public task state $x_q^{\mathrm{pub}}$, ChainSleuth receives the task
description, candidate set, response schema, and declarations of the
allowed and required tools.
% It may retrieve exact transaction evidence before producing a structured
% task-specific answer.

During execution, ChainSleuth follows a bounded interaction loop.
At iteration $j$, given the public task input and interaction history $H_j$, the model either issues an allowed tool call
\[
a_j=\pi_\theta(H_j,x_q^{\mathrm{pub}}),
\]
receives a sanitized observation
\[
o_j=\mathcal{T}(a_j),
\]
and updates
\[
H_{j+1}=H_j\cup\{a_j,o_j\},
\]
or terminates with a structured answer.
Execution stops when the answer is produced, the evidence coverage is reached, no further progress is made, or the tool budget is exhausted.
% Tool results preserve the queried relation, execution status, and sufficient transaction evidence for auditing, while removing all case labels and oracle annotations.

% DTE uses exact neighbor and edge queries to verify the requested directed
% relation and return its existence, transaction count, edge identifiers, and
% transaction hashes.
For tasks \textit{Action}, \textit{Path}, and \textit{Subgraph}, ChainSleuth retrieves transaction-level evidence for public candidate relations and then predicts an action distribution
\[
\pi_\theta(u\mid s_t,e),
\qquad
u\in\mathcal{U},
\]
for each candidate relation $e$ at state $s_t$.
Only relations assigned \texttt{follow} with probability above the declared threshold may be expanded.
Task \textit{Action} performs this decision once, \textit{Path} repeatedly applies it along a bounded trajectory, and \textit{Subgraph} first estimates a seed distribution $\rho_\theta(v\mid x_q^{\mathrm{pub}})$ before applying the same edge policy.

% To control evidence access and improve reproducibility, ChainSleuth may query only relations contained in the public candidate universe, and duplicate, out-of-universe, or over-budget calls are rejected before execution.
After evidence collection terminates, a final synthesis step produces the required JSON response with tools disabled.
% The complete sanitized tool trace is retained for protocol and grounding evaluation, but hidden chain-of-thought is neither stored nor evaluated.

% The context-only ablation uses the same task semantics and structured output
% schema, but receives a fixed public graph snapshot and performs no tool calls.
% Its comparison with the tool-grounded profile isolates the contribution of
% active evidence retrieval from reasoning over pre-supplied context.

\section{Experiments}

We evaluate CryptoFlowBench at two levels.
Task \textit{Trans} and \textit{Action} assess fundamental investigation capabilities, including exact tool invocation, evidence grounding, and local action selection.
% Because conventional heuristic-based procedures can also be utilized as agent tools, and 
Because these diagnostic tasks do not capture full multi-hop tracing, their results are reported in Appendix.C.
% , the results and discussions of \textit{Trans} and \textit{Action} are placed in the Appendix.A.
Task \textit{Path} and \textit{Subgraph} constitute the core tracing benchmark, evaluating sequential investigation under cumulative error and seedless suspicious-region discovery, respectively.
For these tasks, we compare ChainSleuth with task-appropriate heuristic- and learning-based baselines under matched candidate nodes, case splits, and inference budgets.

\subsection{Experimental Settings}

\begin{table}[t]
    \centering
    \small
    \caption{Task distribution by incident.}
    \label{tab:case_task_distribution}
    \scalebox{0.8}{
        \input{tables/task_distribution.tex}
    }
\end{table}

\paragraph{Dataset and Splits}

We construct the benchmark from 22 publicly documented Ethereum incidents using the report collection, on-chain alignment, and deterministic task generation pipeline described above.
Table~\ref{tab:case_task_distribution} reports their distribution across incidents.
To prevent case-specific information leakage, all learning-based methods are trained, validated, and tested using case-disjoint splits.
% No transaction, address, path annotation, or derived label from a target case
% is available during training or threshold selection.
% All methods are evaluated on the same public candidate universe and task-specific
% resource budgets.

\paragraph{Agent Models}

We evaluate ChainSleuth using \textsc{Qwen3.6-27B}~\cite{qwen3.6-27b}, \textsc{DeepSeek-V4-Flash} (DS-V4-Flash), and \textsc{DeepSeek-V4-Pro} (DS-V4-Pro)~\cite{deepseekai2026deepseekv4} as LLM.
% To fully utilize the advantages of tool usage, we also wrap detectors as tools, whose result is reported standalone.

% \paragraph{Evaluation Profiles}

% The primary LLM-agent result uses the \emph{tool-grounded} profile.
% A separately generated \emph{context-only} is reported only with prefix ``Context-''.

\paragraph{Baselines}

We compare ChainSleuth with three groups of task-appropriate baselines. 
We group heuristic baselines into predefined rule-based and graph-based tracing methods.

\textbf{Rule-based tracing.}
We adapt Poison, Haircut, and FIFO~\cite{moser2014towards}, together with LIFO and TIHO~\cite{tironsakkul2019probing}, to the Ethereum account-based setting.
Taint is propagated separately by asset, and all methods use the same initial taint budget, candidate universe, and rollout limits.

\textbf{Graph-based tracing.}
We include TRacer~\cite{wu2023tracer} and adaptations inspired by DenseFlow and DenseFlow+~\cite{lin2024denseflow,wu5076472shadows}.
Because these methods are not originally designed for the present action interface, we map their graph scores to candidate-edge rankings and follow decisions without access to target-case labels.

\textbf{Learning-based classifiers.}
We train logistic regression (LR), random forest (RF), multilayer perceptron (MLP), GCN~\cite{weber2019anti}, and XGBoost~\cite{elmougy2023demystifying} using public candidate-edge features and case-disjoint training folds.
We also utilize several blockchain oriented GNN-based methods, including I$^2$BGNN~\cite{shen2021identity}, PEAE-GNN~\cite{huang2024peae}, and TokenScout~\cite{wu2024tokenscout}.
All GNN-based methods are adapted to edge-level prediction because their original formulations classify transaction nodes rather than candidate address relations.

A frozen XGBoost node classifier is shared across all edge methods for predicted-seed evaluation in the \textit{Subgraph} task, the selection of node classifier will be discussed below.

% \paragraph{Action Mapping}

% The primary common comparison is binary edge relevance:
% \emph{follow} versus \emph{non-follow}.
% For methods that naturally produce only a scalar edge score, the score is used
% to rank candidate relations and select follow edges under a validation-fixed
% threshold.
% Such methods are not assigned artificial \emph{inspect}, \emph{stop}, or
% \emph{ignore} labels.
% Four-way action accuracy is reported only for methods that explicitly output
% the complete action distribution
% $\{\mathtt{follow},\mathtt{inspect},\mathtt{stop},\mathtt{ignore}\}$.

\paragraph{Evaluation Metrics}

All results use deterministic primary scorers and are macro-averaged across incidents. 
We report 95\% confidence intervals in Appendix~D.
For the \textit{Path} task, complete-path rate (Comp.) and Path F1 evaluate trajectory reconstruction, teacher-forced accuracy (TF Acc.) isolates local decision quality without accumulated rollout error, and over-expansion (Over-exp.) measures continuation beyond the verified boundary.
For the \textit{Subgraph} task, Path F1 using predicted seeds (Pred F1) evaluates the complete pipeline of seed identification and suspicious-region expansion, while Path F1 using oracle seeds (Oracle F1) removes seed identification errors and isolates propagation quality; path precision (Path P) and recall (Path R) further characterize the trade-off between noise suppression and suspicious-region coverage.
For ChainSleuth, structured-output validity and tool-protocol pass rate are also reported to distinguish tracing performance from execution and protocol failures.

\subsection{Comparison Experiments}

\input{tables/table_macros}
\begin{table}[t!]
    \centering
    \small
    \caption{\textit{Path} results on 51 tasks from 16 cases. $^*$ indicates ChainSleuth and $^+$ indicates detector tool enabled.}
    \label{tab:pc_path_quality}
    \scalebox{0.9}{
        \input{tables/pc_result}

    }
\end{table}

\begin{table}[t!]
    \centering
    \small
    \caption{\textit{Subgraph} results on the valid-seed, positive-path subset of 22 tasks from 22 cases. $^*$ indicates ChainSleuth and $^+$ indicates detector tool enabled.}
    \label{tab:snf_valid_seed_quality}
    \scalebox{0.9}{
        \input{tables/snf_valid_seed_result}

    }
\end{table}

\begin{table}[t!]
    \centering
    \small
    \caption{Execution reliability of ChainSleuth on path completion. All the methods are ChainSleuth and $^+$ indicates detector tool enabled.}
    \label{tab:agent_pc_reliability}
    \scalebox{0.9}{
        \input{tables/agent_pc_reliability}

    }
\end{table}

Table~\ref{tab:pc_path_quality} and~\ref{tab:snf_valid_seed_quality} show the result of baselines and ChainSleuth on the \textit{Path} and \textit{Subgraph} tasks, respectively.

\paragraph{ChainSleuth versus Specialized Tracing Methods}

ChainSleuth shows its clearest advantage on the practical tracing objectives that require multi-step coordination and control of irrelevant expansion.
On the \textit{Path}, Qwen3.6-27B achieves a complete-path rate of 0.665 and a Path F1 of 0.734, substantially outperforming the strongest conventional baselines.
It also reduces over-expansion to 0.116, compared with 0.158 for the strongest baseline, indicating that ChainSleuth can revise local hypotheses and terminate irrelevant branches before their errors accumulate along the trajectory.
% Although learned models remain competitive in teacher-forced accuracy, with MLP reaching 0.878, their free-rollout performance is considerably weaker, suggesting that accurate isolated decisions do not necessarily translate into reliable multi-step tracing.
A similar pattern appears on the \textit{Subgraph}, where Qwen3.6-27B obtains the highest predicted-seed F1 of 0.375 and the highest path recall of 0.568.
RF performs better when an oracle seed is supplied, achieving an oracle-seed F1 of 0.588 and a path precision of 0.553, but its predicted-seed F1 decreases to 0.309.
These results indicate that specialized classifiers remain strong when the starting point is known, whereas ChainSleuth is more effective in the end-to-end setting that jointly requires seed discovery, evidence acquisition, and subsequent path expansion.

% \paragraph{Tool-Grounded versus Materialized Context}

% Agent-Qwen and Context-Qwen achieve comparable tracing quality despite receiving evidence through different mechanisms.
% On PC, the tool-grounded agent reaches a Path F1 of 0.734, close to the 0.755 obtained with the complete materialized incident context, while both profiles achieve the same teacher-forced accuracy of 0.824.
% Context-Qwen performs somewhat better in complete-path recovery and over-expansion, reaching 0.738 and 0.075 compared with 0.665 and 0.116 for Agent-Qwen, which reflects the advantage of receiving a well-organized incident graph in advance.
% On SNF, however, Agent-Qwen achieves a higher predicted-seed F1 and path recall than Context-Qwen, with scores of 0.375 versus 0.308 and 0.568 versus 0.391, respectively.
% The two profiles therefore exhibit different precision--coverage trade-offs rather than a uniform accuracy advantage for either setting.
% Overall, the results show that the agent can recover sufficient task-relevant evidence through a bounded number of tool calls to approach the decision quality obtained from a pre-materialized incident context, while retaining explicit access to fine-grained transactions and their provenance.

\paragraph{Effect of LLM Backends}

The performance gap between Qwen3.6-27B and the DeepSeek-V4-Flash/Pro backends is closely associated with execution reliability rather than being attributable solely to their underlying reasoning capability.
As shown in Table~\ref{tab:agent_pc_reliability}, on the \textit{Path} task, Qwen3.6-27B achieves a structured-valid rate of 1.000 and a tool-protocol pass rate of 0.979, whereas DeepSeek-V4-Flash reaches only 0.318 and 0.099, respectively.
% These failures prevent DeepSeek-V4-Flash from consistently issuing valid tool calls, acquiring external evidence, and returning evaluable tracing trajectories, resulting in a Path F1 of only 0.202.
% DeepSeek-V4-Pro improves structured validity to 0.780 and tool-protocol compliance to 0.292, accompanied by a substantial increase in Path F1 to 0.541 and complete-path rate to 0.569.
DeepSeek-V4-Pro improves structured validity to 0.780 and tool-protocol compliance to 0.292, which is accompanied by substantially better tracing performance.
The same ordering is observed on \textit{Subgraph}, where the predicted-seed F1 rises from 0.094 for DeepSeek-V4-Flash to 0.213 for DeepSeek-V4-Pro, but remains below the 0.375 achieved by Qwen3.6-27B.
The consistent relationship between protocol reliability and downstream tracing performance suggests that execution reliability is an important contributor to the observed backend gap.

\paragraph{Detector as Tool}

We also evaluate the performance of ChainSleuth with detector as tool, as show in Table~\ref{tab:pc_path_quality} and Table~\ref{tab:snf_valid_seed_quality}.
In detail, we utilize LR, RF, and XGBoost as tools, as those methods can perform a decent result in the \textit{Action}, \textit{Path}, and \textit{Subgraph} tasks, comparing in baselines.
Adding the detector tool increases \textit{Path} completeness from 0.665 to 0.760, but decreases Path F1 from 0.734 to 0.628 and increases over-expansion from 0.116 to 0.221.
On the \textit{Subgraph} task, it degrades all reported metrics, suggesting that detector scores may bias the agent toward locally plausible but globally inconsistent expansions.

\subsection{Seed-Detector Sensitivity}

\begin{table}[t]
    \centering
    \small
    \caption{Case-disjoint comparison of candidate seed detectors under different illicit-node ratios in the training set.}
    \label{tab:seed_detector_sensitivity}
    \setlength{\tabcolsep}{4.5pt}
    \input{tables/seed_detector_sense}
\end{table}

To ensure a fair comparison in \textit{Subgraph}, all baselines share the same case-disjoint seed detector under identical public features and target-case splits.
As shown in Table~\ref{tab:seed_detector_sensitivity}, RF and XGBoost achieve comparable ranking performance and both outperform LR, indicating that the selected detector is not an unusually favorable choice.
We therefore use XGBoost for predicted-seed evaluation and report oracle-seed results separately to isolate propagation performance.

\section{Conclusion}

We introduce a case-grounded benchmark for Ethereum fund-flow tracing that unifies transaction verification, local decisions, multi-step path completion, and seedless subgraph filtering under strict public--oracle separation.
We further develop ChainSleuth that retrieves auditable on-chain evidence and makes structured tracing decisions without requiring the full transaction graph in context.
Experiments show that conventional methods remain competitive on single-step decisions, whereas ChainSleuth performs better on sequential and seed-uncertain tracing.
These results highlight the potential of evidence-grounded agents for practical and reproducible cryptocurrency investigations.

\textbf{Social Impact}.
Cryptocurrency-enabled crime causes substantial financial harm, yet practical fund tracing remains difficult due to pseudonymous accounts and complex multi-hop transactions. 
We provide a new AI perspective by modeling tracing as a case-grounded, sequential investigation problem rather than static illicit-activity detection. 
CryptoFlowBench supports reproducible evaluation across factual verification, local decisions, path reconstruction, and seedless discovery, while ChainSleuth illustrates how tool-augmented agents can acquire evidence, revise earlier decisions, and limit irrelevant expansion. 
This work provides a clear foundation for follow-up research and future decision-support tools for cryptocurrency investigators.

% \section{Acknowledgments}

\bibliography{main}

\end{document}

%% file: tables/task_info.tex
% \begin{tabular}{lll}
%     \hline
%     Case & Type & Primary source \\
%     \hline
%     01 Ronin & \multirow{9}{*}{\shortstack{Credential/\\UI compromise}} & U.S. Treasury \\
%     03 Harmony & & Harmony \\
%     09 CoinEx & & CoinEx \\
%     10 HTX HECO Bridge & & HTX \\
%     11 Stake.com & & Stake.com \\
%     12 WazirX & & WazirX \\
%     13 Radiant Capital & & Radiant Capital \\
%     14 PlayDapp & & PlayDapp \\
%     15 Bybit & & Bybit \\
%     \hline
%     02 Wormhole & \multirow{10}{*}{Protocol exploit} & CertiK \\
%     04 Nomad Bridge & & Nomad \\
%     05 BNB Bridge & & BNB Chain \\
%     06 Euler Finance & & Euler Finance \\
%     07 Curve Vyper Pools & & Vyper \\
%     08 KyberSwap Elastic & & KyberSwap \\
%     16 Cetus Protocol & & Sui Foundation \\
%     17 Prisma Finance & & Prisma Finance \\
%     19 Penpie & & Penpie \\
%     21 Dough Finance & & Dough Finance \\
%     \hline
%     18 UwU Lend & Oracle manipulation & SlowMist \\
%     20 Munchables & Backdoor & Halborn \\
%     22 Atomic Wallet & Wallet compromise & Atomic Wallet \\
%     \hline
% \end{tabular}

\begin{tabular}{llc}
    \hline
    \textbf{Case} & \textbf{Primary source} & \textbf{Type} \\
    \hline
    01 Ronin & U.S. Treasury & \multirow{9}{*}{\shortstack{Credential/\\UI compromise}} \\
    02 Harmony & Harmony & \\
    03 CoinEx & CoinEx & \\
    04 HTX HECO Bridge & HTX & \\
    05 Stake.com & Stake.com & \\
    06 WazirX & WazirX & \\
    07 Radiant Capital & Radiant Capital & \\
    08 PlayDapp & PlayDapp & \\
    09 Bybit & Bybit & \\
    \hline
    10 Wormhole & CertiK & \multirow{10}{*}{Protocol exploit} \\
    11 Nomad Bridge & Nomad & \\
    12 BNB Bridge & BNB Chain & \\
    13 Euler Finance & Euler Finance & \\
    14 Curve Vyper Pools & Vyper & \\
    15 KyberSwap Elastic & KyberSwap & \\
    16 Cetus Protocol & Sui Foundation & \\
    17 Prisma Finance & Prisma Finance & \\
    18 Penpie & Penpie & \\
    19 Dough Finance & Dough Finance & \\
    \hline
    20 UwU Lend & SlowMist & Oracle manipulation \\
    \hline
    21 Munchables & Halborn & Backdoor \\
    \hline
    22 Atomic Wallet & Atomic Wallet & Wallet compromise \\
    \hline
\end{tabular}

%% file: tables/task_distribution.tex
\begin{tabular}{lrrrrr}
    \hline
    \textbf{Case} & \textbf{Trans} & \textbf{Action} & \textbf{Path} & \textbf{Subgraph} & \textbf{Total} \\
    \hline
    01 Ronin & 658 & 11 & 7 & 13 & 689 \\
    02 Harmony & 1,720 & 4 & 1 & 25 & 1,750 \\
    03 CoinEx & 1,216 & 4 & 0 & 21 & 1,241 \\
    04 HTX HECO Bridge & 220 & 14 & 5 & 13 & 252 \\
    05 Stake.com & 148 & 10 & 2 & 13 & 173 \\
    06 WazirX & 706 & 5 & 1 & 25 & 737 \\
    07 Radiant Capital & 1,228 & 7 & 4 & 13 & 1,252 \\
    08 PlayDapp & 1,688 & 5 & 1 & 25 & 1,719 \\
    09 Bybit & 19,584 & 54 & 4 & 25 & 19,667 \\
    10 Wormhole & 202 & 1 & 0 & 13 & 216 \\
    11 Nomad Bridge & 522 & 5 & 1 & 17 & 545 \\
    12 BNB Bridge & 1,594 & 1 & 0 & 13 & 1,608 \\
    13 Euler Finance & 1,504 & 13 & 8 & 25 & 1,550 \\
    14 Curve Vyper Pools & 440 & 8 & 3 & 15 & 466 \\
    15 KyberSwap Elastic & 1,844 & 3 & 0 & 25 & 1,872 \\
    16 Cetus Protocol & 662 & 3 & 0 & 13 & 678 \\
    17 Prisma Finance & 466 & 6 & 3 & 25 & 500 \\
    18 Penpie & 328 & 5 & 3 & 13 & 349 \\
    19 Dough Finance & 1,610 & 5 & 2 & 25 & 1,642 \\
    20 UwU Lend & 1,620 & 3 & 0 & 13 & 1,636 \\
    21 Munchables & 238 & 7 & 2 & 13 & 260 \\
    22 Atomic Wallet & 350 & 14 & 4 & 14 & 382 \\
    \hline
    Total & 38,548 & 188 & 51 & 397 & 39,184 \\
    Share (\%) & 98.4 & 0.5 & 0.1 & 1.0 & 100.0 \\
    \hline
\end{tabular}

%% file: tables/table_macros.tex
% Shared compact confidence-interval formatting: mean, lower, upper.
\newcommand{\mci}[3]{\ensuremath{#1}}
\newcommand{\bestmci}[3]{\ensuremath{\mathbf{#1}}}
\newcommand{\secmci}[3]{\ensuremath{\underline{#1}}}

%% file: tables/pc_result.tex
\begin{tabular}{lcccc}
\hline
\textbf{Method} & Comp. $\uparrow$ & Path F1 $\uparrow$ & TF Acc. $\uparrow$ & Over-exp. $\downarrow$ \\
\hline
Poison & \mci{0.330}{0.188}{0.480} & \mci{0.345}{0.243}{0.443} & \mci{0.396}{0.309}{0.500} & \mci{0.644}{0.515}{0.762} \\
Haircut & \mci{0.330}{0.186}{0.484} & \mci{0.311}{0.218}{0.401} & \mci{0.396}{0.308}{0.504} & \mci{0.674}{0.547}{0.786} \\
FIFO & \mci{0.422}{0.243}{0.609} & \mci{0.340}{0.232}{0.456} & \mci{0.560}{0.491}{0.640} & \mci{0.628}{0.487}{0.757} \\
LIFO & \mci{0.301}{0.121}{0.495} & \mci{0.328}{0.206}{0.449} & \mci{0.476}{0.379}{0.574} & \mci{0.651}{0.506}{0.791} \\
TIHO & \mci{0.313}{0.158}{0.479} & \mci{0.307}{0.205}{0.412} & \mci{0.604}{0.503}{0.693} & \mci{0.646}{0.507}{0.774} \\
TRacer & \mci{0.087}{0.000}{0.237} & \mci{0.186}{0.082}{0.304} & \mci{0.643}{0.580}{0.710} & \mci{0.158}{0.076}{0.242} \\
DenseFlow+ & \mci{0.296}{0.131}{0.473} & \mci{0.315}{0.154}{0.483} & \mci{0.517}{0.399}{0.631} & \mci{0.332}{0.186}{0.484} \\
\hline
LR & \mci{0.426}{0.262}{0.601} & \mci{0.432}{0.314}{0.566} & \mci{0.718}{0.647}{0.792} & \mci{0.590}{0.464}{0.705} \\
RF & \mci{0.374}{0.212}{0.558} & \mci{0.379}{0.239}{0.530} & \mci{0.813}{0.753}{0.873} & \mci{0.603}{0.442}{0.753} \\
MLP & \mci{0.253}{0.113}{0.426} & \mci{0.212}{0.112}{0.328} & \bestmci{0.878}{0.830}{0.925} & \mci{0.270}{0.124}{0.442} \\
GCN & \mci{0.336}{0.190}{0.495} & \mci{0.300}{0.204}{0.398} & \mci{0.774}{0.690}{0.852} & \mci{0.605}{0.471}{0.733} \\
I²BGNN & \mci{0.305}{0.161}{0.459} & \mci{0.279}{0.186}{0.372} & \mci{0.623}{0.496}{0.741} & \mci{0.450}{0.314}{0.593} \\
PEAE-GNN & \mci{0.290}{0.141}{0.452} & \mci{0.267}{0.168}{0.368} & \mci{0.572}{0.445}{0.698} & \mci{0.409}{0.255}{0.572} \\
TokenScout & \mci{0.307}{0.164}{0.461} & \mci{0.286}{0.188}{0.383} & \mci{0.659}{0.548}{0.768} & \mci{0.414}{0.266}{0.571} \\
XGB & \mci{0.276}{0.132}{0.443} & \mci{0.279}{0.156}{0.421} & \secmci{0.854}{0.796}{0.910} & \mci{0.325}{0.175}{0.484} \\
\hline
\rowcolor{black!10}
DS-V4-Flash $^*$ & \mci{0.092}{0.030}{0.168} & \mci{0.202}{0.110}{0.304} & \mci{0.244}{0.126}{0.382} & \bestmci{0.097}{0.019}{0.191} \\
\rowcolor{black!10}
DS-V4-Pro $^*$ & \mci{0.569}{0.388}{0.744} & \mci{0.541}{0.426}{0.652} & \mci{0.570}{0.442}{0.696} & \mci{0.185}{0.113}{0.268} \\
\rowcolor{black!10}
Qwen3.6-27B $^*$ & \secmci{0.665}{0.448}{0.859} & \bestmci{0.734}{0.662}{0.801} & \mci{0.824}{0.760}{0.881} & \secmci{0.116}{0.042}{0.207} \\
\rowcolor{black!10}
Qwen3.6-27B $^{*+}$ & \bestmci{0.760}{0.606}{0.893} & \secmci{0.628}{0.505}{0.740} & \mci{0.603}{0.483}{0.717} & \mci{0.221}{0.125}{0.324} \\
\hline
\end{tabular}

%% file: tables/snf_valid_seed_result.tex
\begin{tabular}{lcccc}
\hline
\textbf{Method} & Pred F1 $\uparrow$ & Oracle F1 $\uparrow$ & Path P $\uparrow$ & Path R $\uparrow$ 
\\
\hline
Poison & \mci{0.199}{0.115}{0.287} & \mci{0.167}{0.087}{0.252} & \mci{0.295}{0.159}{0.439} & \mci{0.180}{0.092}{0.288} \\
Haircut & \mci{0.199}{0.117}{0.285} & \mci{0.167}{0.086}{0.253} & \mci{0.295}{0.159}{0.439} & \mci{0.180}{0.095}{0.282} \\
FIFO & \mci{0.219}{0.132}{0.311} & \mci{0.228}{0.146}{0.313} & \mci{0.318}{0.189}{0.462} & \mci{0.190}{0.105}{0.297} \\
LIFO & \mci{0.217}{0.130}{0.307} & \mci{0.136}{0.073}{0.204} & \mci{0.379}{0.220}{0.545} & \mci{0.164}{0.098}{0.233} \\
TIHO & \mci{0.237}{0.146}{0.330} & \mci{0.228}{0.135}{0.331} & \mci{0.386}{0.235}{0.553} & \mci{0.202}{0.114}{0.307} \\
TRacer & \mci{0.109}{0.043}{0.183} & \mci{0.137}{0.065}{0.218} & \mci{0.220}{0.083}{0.379} & \mci{0.075}{0.030}{0.124} \\
DenseFlow+ & \mci{0.087}{0.021}{0.174} & \mci{0.126}{0.051}{0.211} & \mci{0.167}{0.045}{0.318} & \mci{0.063}{0.015}{0.130} \\
\hline
LR & \mci{0.303}{0.221}{0.385} & \secmci{0.565}{0.482}{0.645} & \mci{0.473}{0.326}{0.629} & \mci{0.283}{0.182}{0.402} \\
RF & \secmci{0.309}{0.201}{0.423} & \bestmci{0.588}{0.507}{0.668} & \bestmci{0.553}{0.371}{0.727} & \mci{0.252}{0.145}{0.376} \\
MLP & \mci{0.196}{0.091}{0.317} & \mci{0.386}{0.274}{0.501} & \mci{0.394}{0.197}{0.591} & \mci{0.145}{0.061}{0.255} \\
GCN & \mci{0.199}{0.117}{0.287} & \mci{0.423}{0.330}{0.509} & \mci{0.295}{0.167}{0.439} & \mci{0.180}{0.094}{0.288} \\
I²BGNN & \mci{0.176}{0.091}{0.267} & \mci{0.350}{0.237}{0.462} & \mci{0.265}{0.129}{0.417} & \mci{0.161}{0.073}{0.267} \\
PEAE-GNN & \mci{0.166}{0.081}{0.260} & \mci{0.383}{0.285}{0.479} & \mci{0.250}{0.114}{0.409} & \mci{0.153}{0.066}{0.263} \\
TokenScout & \mci{0.187}{0.104}{0.276} & \mci{0.400}{0.305}{0.491} & \mci{0.280}{0.144}{0.432} & \mci{0.170}{0.083}{0.274} \\
XGB & \mci{0.218}{0.121}{0.323} & \mci{0.391}{0.295}{0.480} & \secmci{0.485}{0.273}{0.682} & \mci{0.148}{0.079}{0.227} \\
\hline
\rowcolor{black!10}
DS-V4-Flash $^*$ & \mci{0.094}{0.000}{0.211} & \mci{0.094}{0.000}{0.211} & \mci{0.081}{0.000}{0.192} & \mci{0.127}{0.000}{0.273} \\
\rowcolor{black!10}
DS-V4-Pro $^*$ & \mci{0.213}{0.089}{0.356} & \mci{0.204}{0.086}{0.340} & \mci{0.166}{0.063}{0.293} & \secmci{0.381}{0.190}{0.571} \\
\rowcolor{black!10}
Qwen3.6-27B $^*$ & \bestmci{0.375}{0.224}{0.537} & \mci{0.306}{0.167}{0.461} & \mci{0.314}{0.174}{0.469} & \bestmci{0.568}{0.364}{0.773} \\
\rowcolor{black!10}
Qwen3.6-27B $^{*+}$ & \mci{0.182}{0.067}{0.315} & \mci{0.271}{0.130}{0.423} & \mci{0.223}{0.076}{0.390} & \mci{0.183}{0.057}{0.333} \\
\hline
\end{tabular}

%% file: tables/agent_pc_reliability.tex
\begin{tabular}{lcc}
\hline
\textbf{Method} & Structure-valid rate $\uparrow$ & Tool-protocol pass rate $\uparrow$ \\
\hline
DS-V4-Flash & \mci{0.318}{0.167}{0.490} & \mci{0.099}{0.000}{0.245} \\
DS-V4-Pro & \mci{0.780}{0.638}{0.897} & \mci{0.292}{0.135}{0.466} \\
Qwen3.6-27B $^{+}$ & \mci{0.802}{0.651}{0.921} & \mci{0.766}{0.622}{0.885} \\
Qwen3.6-27B & \bestmci{1.000}{1.000}{1.000} & \bestmci{0.979}{0.938}{1.000} \\
\hline
\end{tabular}

%% file: tables/seed_detector_sense.tex
\begin{tabular}{clcccc}
    \hline
    \textbf{Train rate} & \textbf{Method} & AP $\uparrow$ & Hit@1 $\uparrow$ &
    Hit@5 $\uparrow$ & MRR $\uparrow$ \\
    \hline
    \multirow{3}{*}{1\%}
    & LR  & 0.651 & 0.546 & 0.955 & 0.746 \\
    & RF  & 0.881 & \textbf{0.955} & \textbf{1.000} &
    \textbf{0.970} \\
    & XGB & \textbf{0.883} & \textbf{0.955} & \textbf{1.000} &
    \textbf{0.970} \\
    \hline
    \multirow{3}{*}{5\%}
    & LR  & 0.659 & 0.546 & 0.955 & 0.735 \\
    & RF  & 0.862 & \textbf{0.909} & \textbf{1.000} &
    \textbf{0.947} \\
    & XGB & \textbf{0.877} & 0.864 & \textbf{1.000} & 0.921 \\
    \hline
    \multirow{3}{*}{10\%}
    & LR  & 0.688 & 0.500 & \textbf{1.000} & 0.750 \\
    & RF  & 0.827 & \textbf{0.864} & \textbf{1.000} &
    \textbf{0.921} \\
    & XGB & \textbf{0.829} & 0.773 & \textbf{1.000} & 0.886 \\
    \hline
\end{tabular}